\documentclass[a4paper,12pt,onecolumn]{article}
\usepackage[dvips]{graphicx}
\usepackage{txfonts} 

\begin{document}
\begin{flushleft}
\textbf{\LARGE Magnitude systems in old star catalogues}

\bigskip
{\bf Tomoko Fujiwara}$^*$ \& {\bf Hitoshi Yamaoka}\dag\\

\bigskip
$^*$ {\it Department of Physics, Graduate School of Science, Kyoto Sangyo University, Kyoto 603-8555, Japan\\
E-mail: tomochan@cc.kyoto-su.ac.jp}\\
\dag {\it Department of Physics, Faculty of Science, Kyushu University, 4-2-1, Ropponmatsu, Chuo-ku, Fukuoka 810-8560, Japan\\
E-mail: yamaoka@rc.kyushu-u.ac.jp}\\
\end{flushleft}

\smallskip
\noindent
{\bf Abstract} \\
\noindent
The current system of stellar magnitudes first introduced by 
Hipparchus was strictly defined by Norman Robert Pogson in 1856. 
He based his system on Ptolemy's star catalogue `{\it Almagest}', 
recorded in about 137 A.D., and defined the magnitude-intensity 
relationship on a logarithmic scale. 
   
Stellar magnitudes observed with the naked eye recorded in seven 
old star catalogues were analyzed in order to examine the visual 
magnitude systems. 
Despite that psychophysists have proposed that human's sensitivities 
are on a power-law scale, it is shown that the degree of agreement 
is far better for a logarithmic magnitude than a power-law magnitude. 
It is also found that light ratios in each star catalogue nearly 
equal to 2.512, excluding the brightest (1st) and the dimmest 
(6th and dimmer) stars being unsuitable for the examination. 
It means that the visual magnitudes in old star catalogues 
fully agree with Pogson's logarithmic scale. \\

\noindent
{\bf keywords} stars: general -- historical star catalogues -- stellar magnitude system -- visual magnitude estimates -- astronomical photometry\\

\section{Introduction}
The concept of magnitudes was introduced by Hipparchus in c.2 B.C. (cf. Hearnshaw 1996). 
Hipparchus compiled his catalogue of 850 stars with ecliptical coordinates 
and visual magnitudes. 
This work was triggered by the discovery and the observation of a nova 
(not yet explained) in the constellation Scorpius in 134 B.C. 
He started to record the coordinates and magnitudes of fixed stars in order to aid discoveries of such objects, and to record the brightness.
He defined the brightest 20 stars as 1st magnitude, 
Polaris and stars of the Great Dipper in Ursa Major as 2nd magnitude and stars at the observable limit of the naked eye as 6th magnitude. 
The work of Hipparchus was lost over the years, however, Hipparchus' magnitude system came down through subsequent star catalogues (`{\it Almagest}' etc.).

In the nineteenth century, astronomers tried to define the magnitude 
system more precisely and quantitatively, based on simple arbitrary visual estimates.
Many astronomers (W. and J. Herschel etc.) had already investigated the magnitude-intensity relationship and deduced the logarithmic form. 
Based on Ptolemy's star catalogue `{\it Almagest}', Pogson (1856) proposed adopting a light ratio {\it R} 2.512 for two stars that differ in brightness by one magnitude, defining the magnitude as
   
   \begin{equation}
   m = - \frac{1}{\log{R}} \log{I}. 
   \end{equation}

In the case of $R = 2.512$, this formula could be transformed into 

   \begin{equation}
   m = - 2.5 \log{I}. 
   \end{equation}

This definition is well-known as Pogson scale and is still used in stellar photometry. 
   
In the 1960s, psychophysists propounded that the response of human's sensitivity would be a power law (1961). 
Referring this theory, Schulman \& Cox (1997) suggested that visual magnitude estimates were much better fit to a power law.
Equally, the eye's response to light is a power law, and therefore 
visual magnitude estimates disagreed with the logarithmic system. 
   
Independently, Hearnshaw (1996, 1999) examined `{\it Almagest}', and showed that the magnitudes fitted to the logarithmic scale. 
The light ratio of `{\it Almagest}' is, however, derived as 3.42 being far larger than that of Pogson's formula. 

In order to verify that visual magnitude estimates fit either a logarithm or a power law, we intend to investigate the magnitude systems in old star catalogues. 
In all of the star catalogues mentioned below, stellar magnitudes were estimated with the naked eye and were classified by 1st to 6th based on the Hipparchus' system. 
   
   \begin{enumerate}
     \item {\it Almagest} (Ptolemaios AD127--141)
     \item {\it \d{S}uwar al-Kaw\={a}kib} (al-\d{S}\={u}f\={\i} 986)
     \item {\it Ulugh Beg's Catalogue of stars} (1437)
     \item {\it Astronomiae Instauratae Progymnasmata} (Brahe 1602)
     \item {\it Uranometria} (Bayer 1603)
     \item {\it Historia Coelestis Britannica} (Flamsteed 1725)
     \item {\it Uranometria Nova} (Argelander 1843)
   \end{enumerate}

We refer to `{\it Sky Catalogue 2000.0}' (Hirshfeld et al. 1991) for modern stellar magnitude data. 

In this paper, we present the results of our study of magnitude systems in old star catalogues. 
Magnitude data and their analysis are found in Sect. 2. 
We present and compare historical magnitude data on the chart with a logarithmic scale and a power-law scale in Sect. 3.1. 
The light ratios {\it R} are described in Sect. 3.2. 
The conclusions are summarized in Sect. 4. 

\section{Data and Analysis}

Before we could use data compiled in these catalogues, we had to check the characteristics of old works and correct the magnitude data (see Fujiwara et al. 2003).

In these old star catalogues, magnitude classes were recorded by numbers (1--6) and plus or minus signs which indicated `a little brighter' or `a little dimmer', respectively. 
To quantify these magnitude descriptions completely, we subtracted or added 0.33 according to the plus or minus sign respectively.
For example, we assigned 2.67 for $3+$ and 3.33 for $3-$.  
   
We omitted unsuitable stars as follow: first, stars presently brighter than 1 mag because in the days when these catalogues were recorded, there was no concept of zero or minus magnitude; second, stars that we could  not identify; third, untreatable double or binary stars; finally, known variables with amplitudes larger than 0.5 mag; $o$ Cet (Mira), $\beta$ Per (Algol), $\delta$ Cep, etc. (for detailed descriptions, see also Fujiwara et al. 2003)  

Consequently, we sampled, in total, 2124 naked-eye stars, and in Table~\ref{tbl-1}, the observational (not published) epoch of every star catalogue is given in Column 2, while the number of stars N is shown in Column 3.

We compared each magnitude system of seven old star catalogues to Pogson's magnitude system. 
First, we investigated magnitude data in old catalogues on the chart based on the logarithmic magnitude scale. 
In Fig.~\ref{fig1}, the present magnitudes of stars recorded in each star catalogue are dotted lines and it is naturally equivalent to Pogson's scale. 
The solid lines indicate linear regressions of old magnitudes.

Secondly, we plotted the same magnitude data on a power-law scale chart (see Fig.~\ref{fig2}).
The function of Schulman \& Cox is given as 
   
   \begin{equation}
   m = 5.5556 ( 2.512^{(-0.5)(6-V)} ) + 0.4444.
   \end{equation}
   
This scale is indicated with dotted lines and power-law regressions are shown with solid lines.

\section{Results and Discussion}
\subsection{logarithm vs. power law}

In this study, we compared the magnitude system based on the logarithmic scale with that on the power-law scale. 
As shown in Fig.~\ref{fig1}, on the logarithmic scale, magnitude data recorded in old catalogues correspond exactly with Pogson's logarithmic scale.

It can be seen from Fig.~\ref{fig2} that the function suggested by Schulman \& Cox does not at all fit to magnitude data in old star catalogues. 
Relative to power-law regressions shown with solid lines, for dimmer magnitudes (3--6), regressed functions fit to the magnitude data, on the contrary, those for brighter magnitudes (1--3) deviate notably. 
We could not say that magnitudes fitted to a power-law system unless the data did not have a bias toward proportions at all points on a power-law scale chart. 

In order to investigate that magnitude data in old star catalogues fit better to which scale, we estimate by chi-square tests. 
In Table~\ref{tbl-2}, we can see reduced chi-squares. 
The catalogue ID (listed in Sect.1) is found in Column 1, the reduced chi-square ${{\chi}_{\nu}}^{2}$ on the logarithmic scale are shown in Column 2, and on the power-law scale the reduced chi-square ${{\chi}_{\nu}}^{2}$ are shown in Column 3. 
If a regressed finction fits to the data, the reduced chi-squar ${{\chi}_{\nu}}^{2}$ should be small. 
   
As shown in Table~\ref{tbl-2}, all reduced chi-squars ${{\chi}_{\nu}}^{2}$ on the logarithmic scale are much small, i.e. estimated linear regressions are almost proper. 
Contrarily, on the power-law scale, each reduced chi-squar is too large. 
It means that regressions on power-law scale could be alterable to another function, and estimated regressions do not fit at all. 
It suggests that historical magnitudes also disagree with a power-law system. 
Magnitude systems in all old star catalogues do not fit to the power-law scale but instead, to the logarithmic scale. 

\subsection{Examine of the light ratio}

\begin{figure}
   \centering
    \begin{tabular}{cc}
      \resizebox{60mm}{!}{\includegraphics{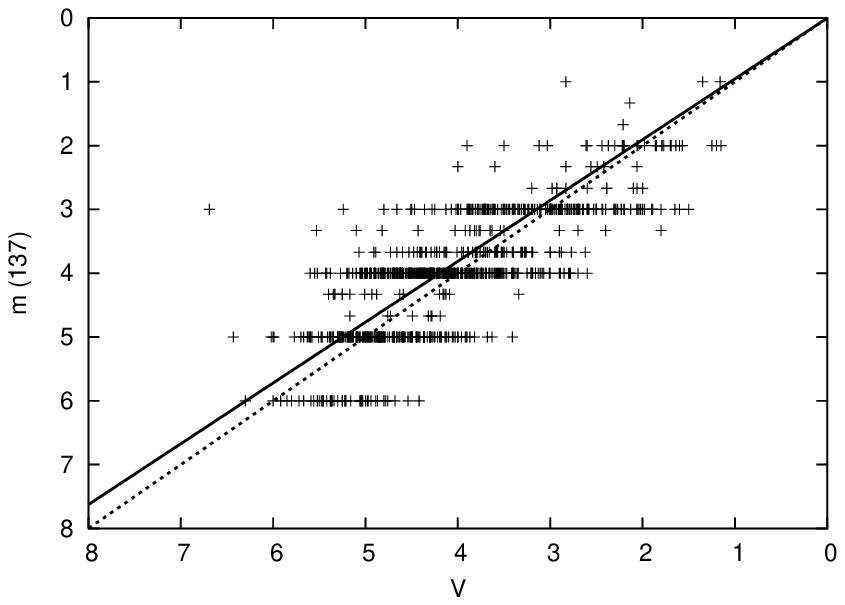}} &
      \resizebox{60mm}{!}{\includegraphics{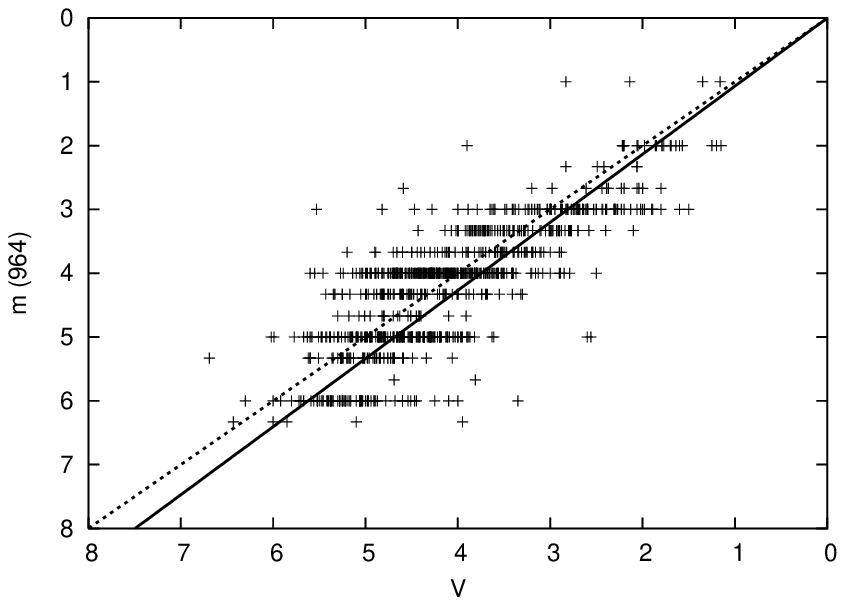}} \\
      \resizebox{60mm}{!}{\includegraphics{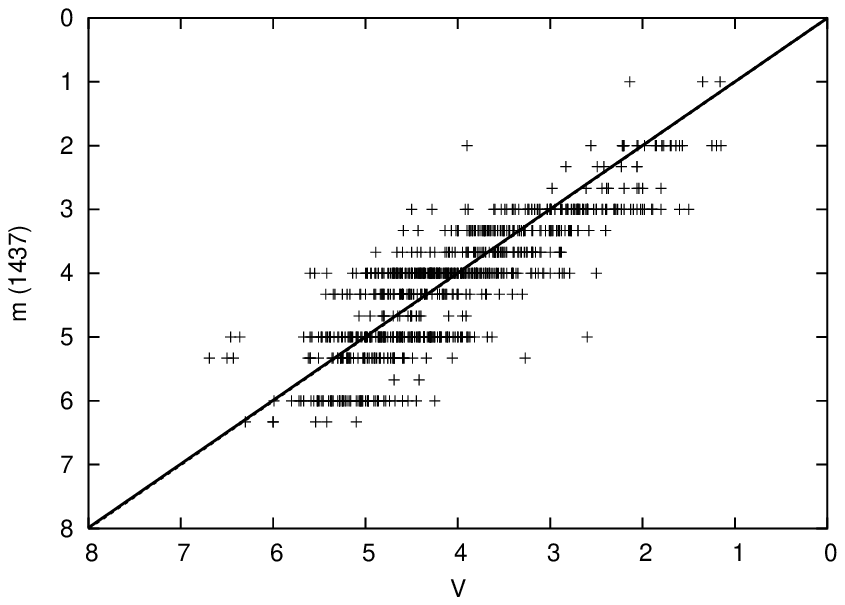}} &
      \resizebox{60mm}{!}{\includegraphics{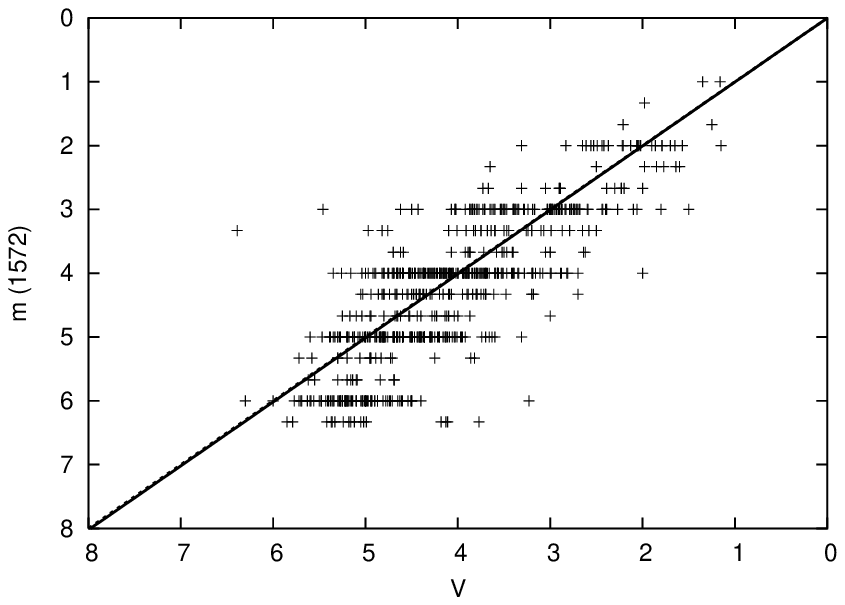}} \\
      \resizebox{60mm}{!}{\includegraphics{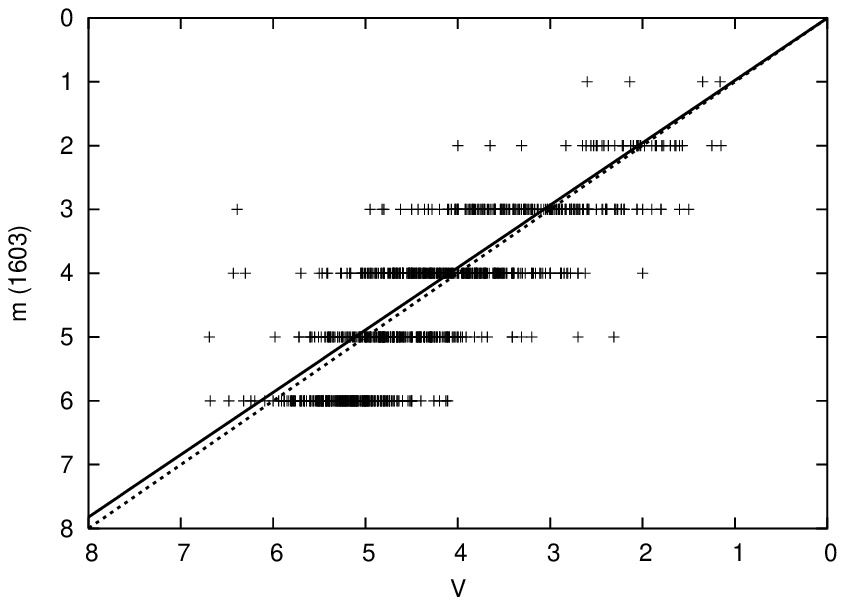}} &
      \resizebox{60mm}{!}{\includegraphics{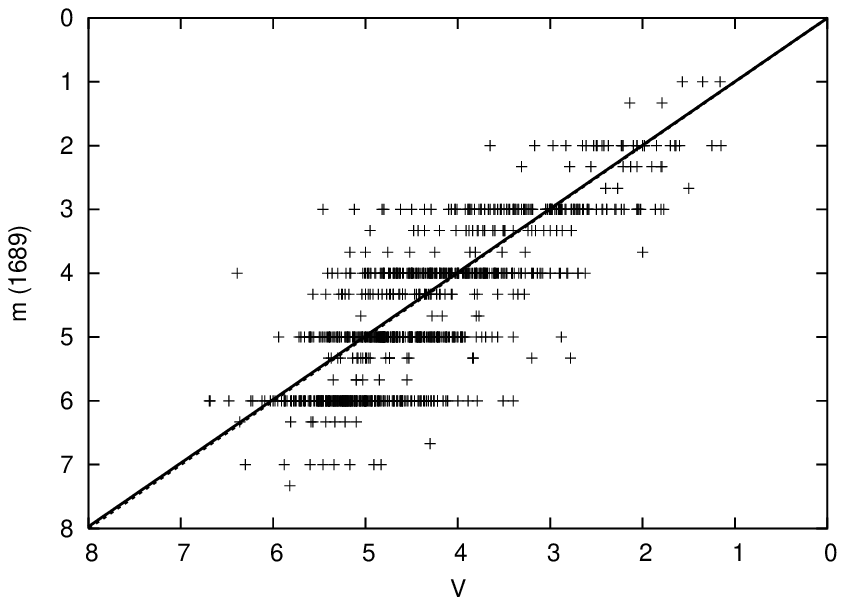}} \\
      \resizebox{60mm}{!}{\includegraphics{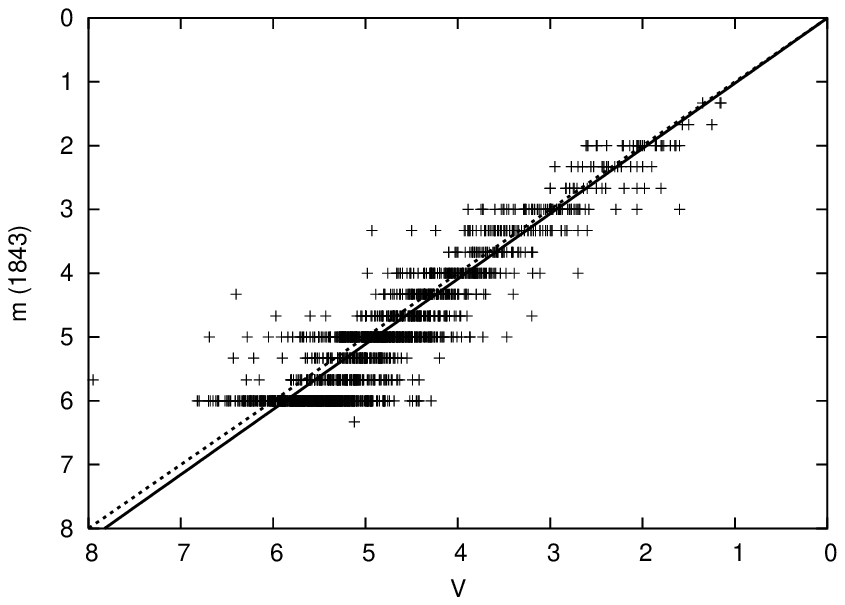}} \\
    \end{tabular}

   \caption{magnitude systems on the logarithmic scale. 
      Dotted lines indicate Pogson's scale and solid lines
      indicate linear regressions. 
      \label{fig1}}
   \end{figure}

\begin{figure}
   \centering
    \begin{tabular}{cc}
      \resizebox{60mm}{!}{\includegraphics{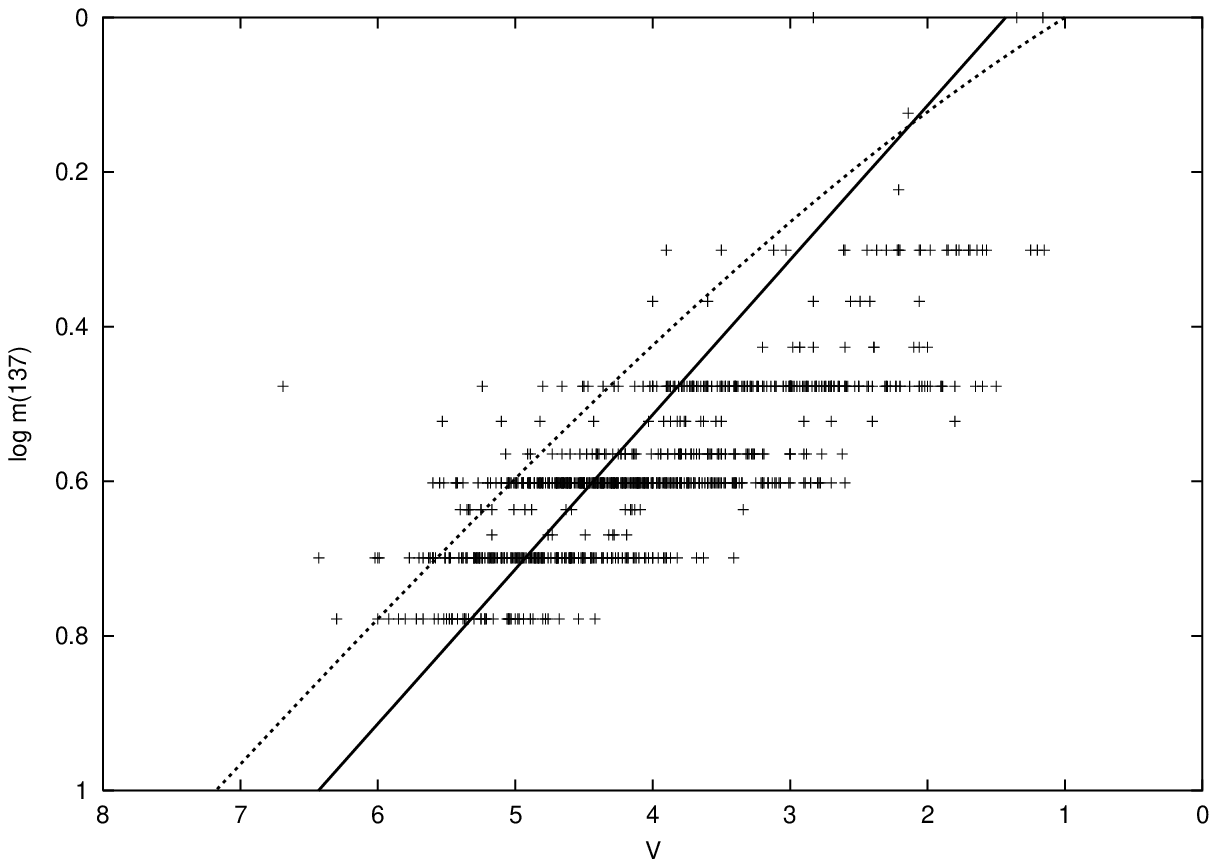}} &
      \resizebox{60mm}{!}{\includegraphics{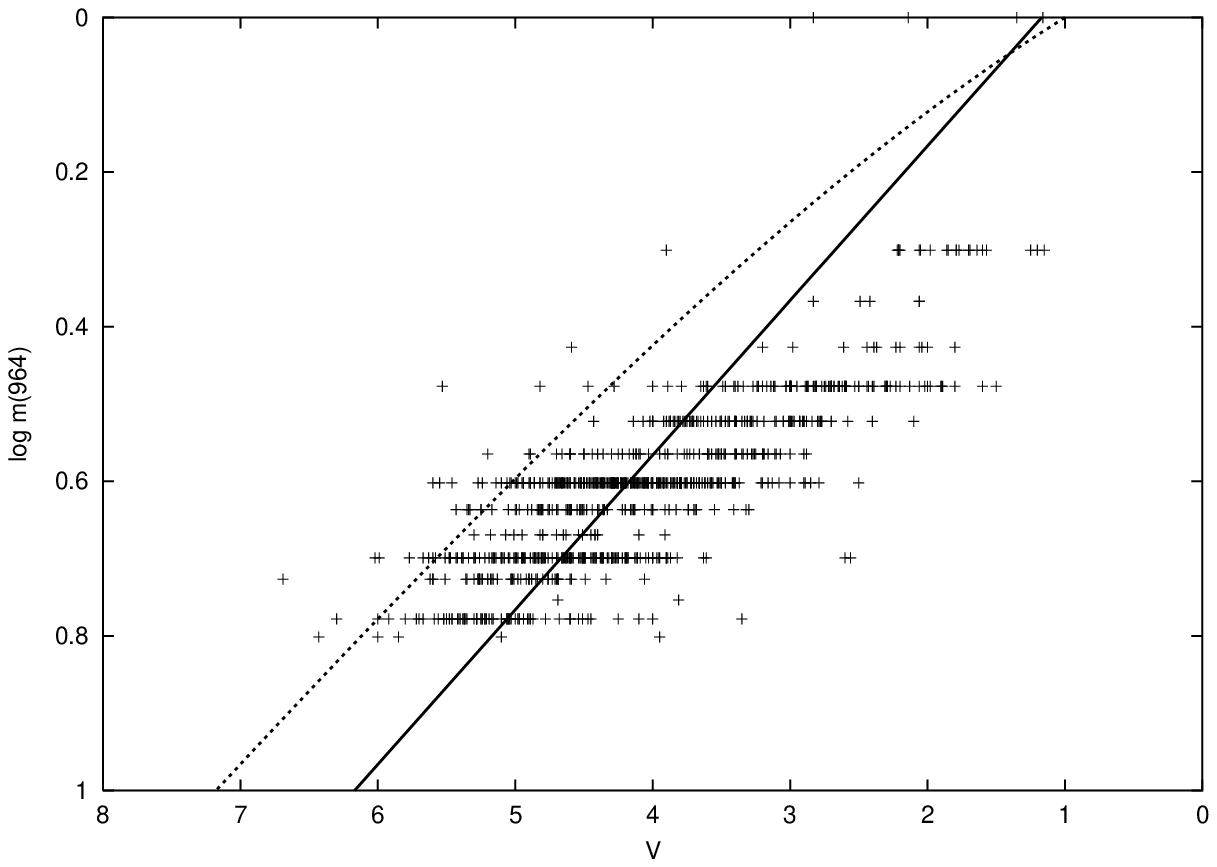}} \\
      \resizebox{60mm}{!}{\includegraphics{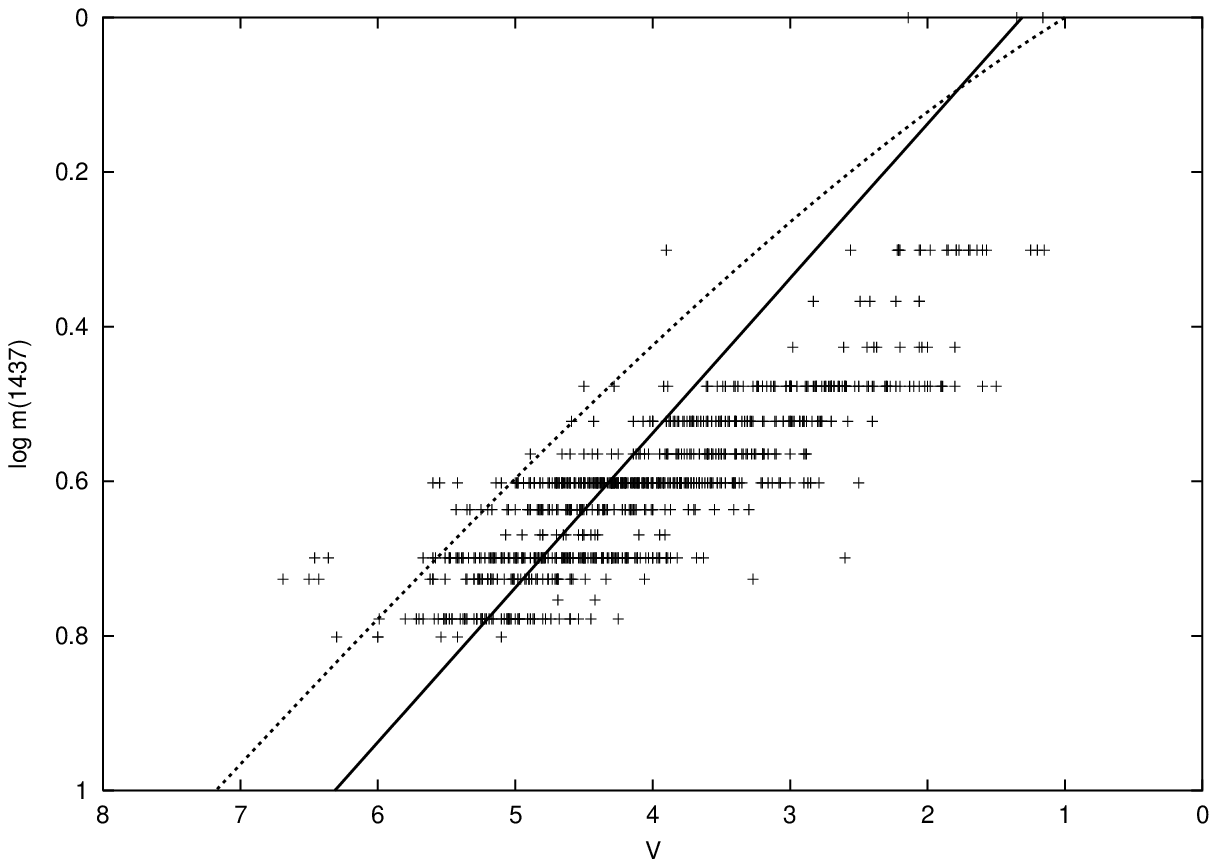}} &
      \resizebox{60mm}{!}{\includegraphics{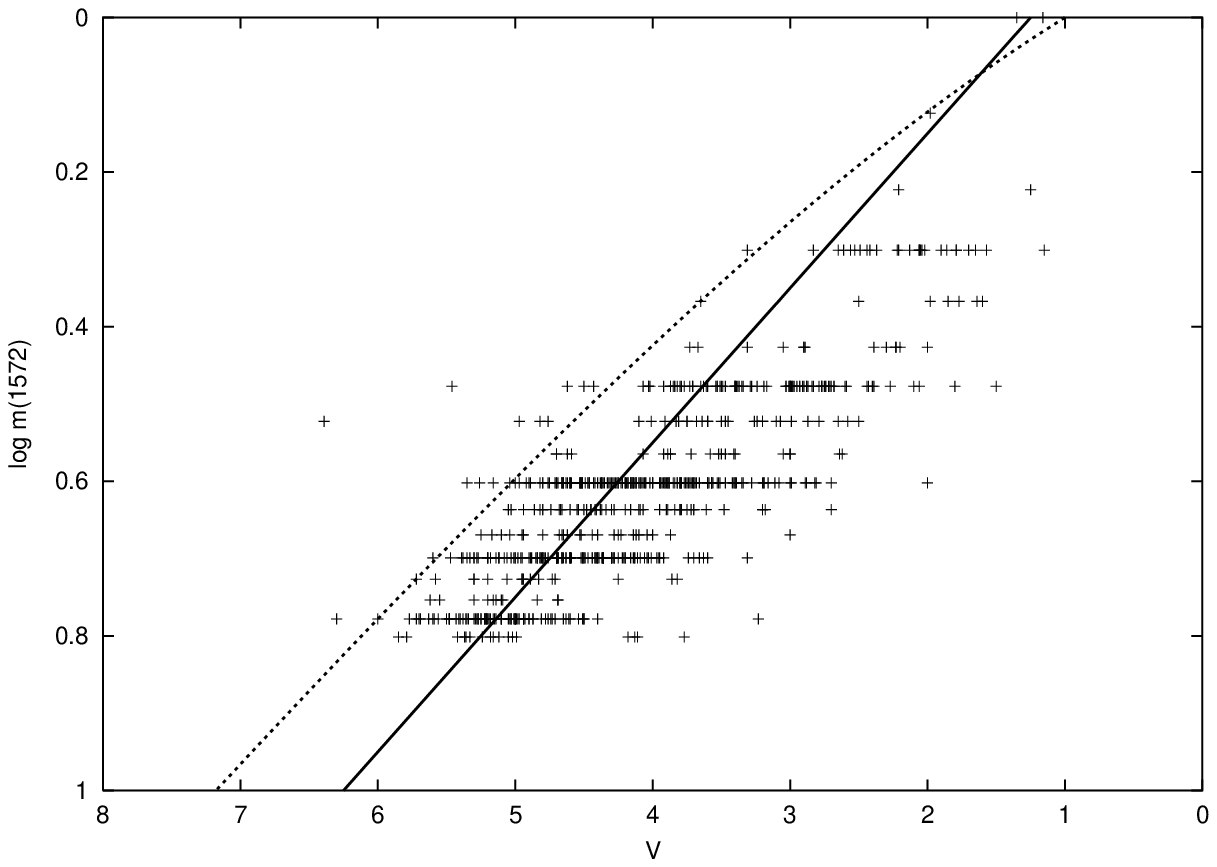}} \\
      \resizebox{60mm}{!}{\includegraphics{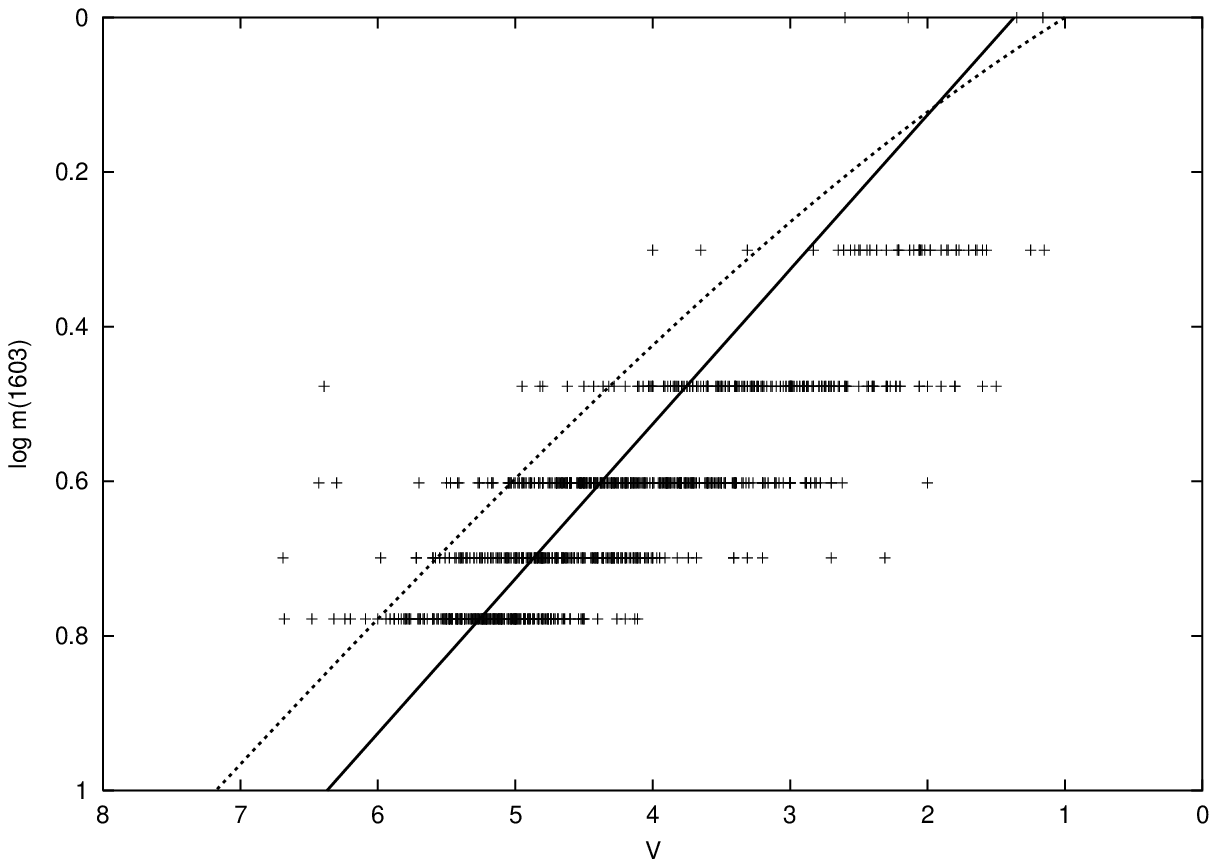}} &
      \resizebox{60mm}{!}{\includegraphics{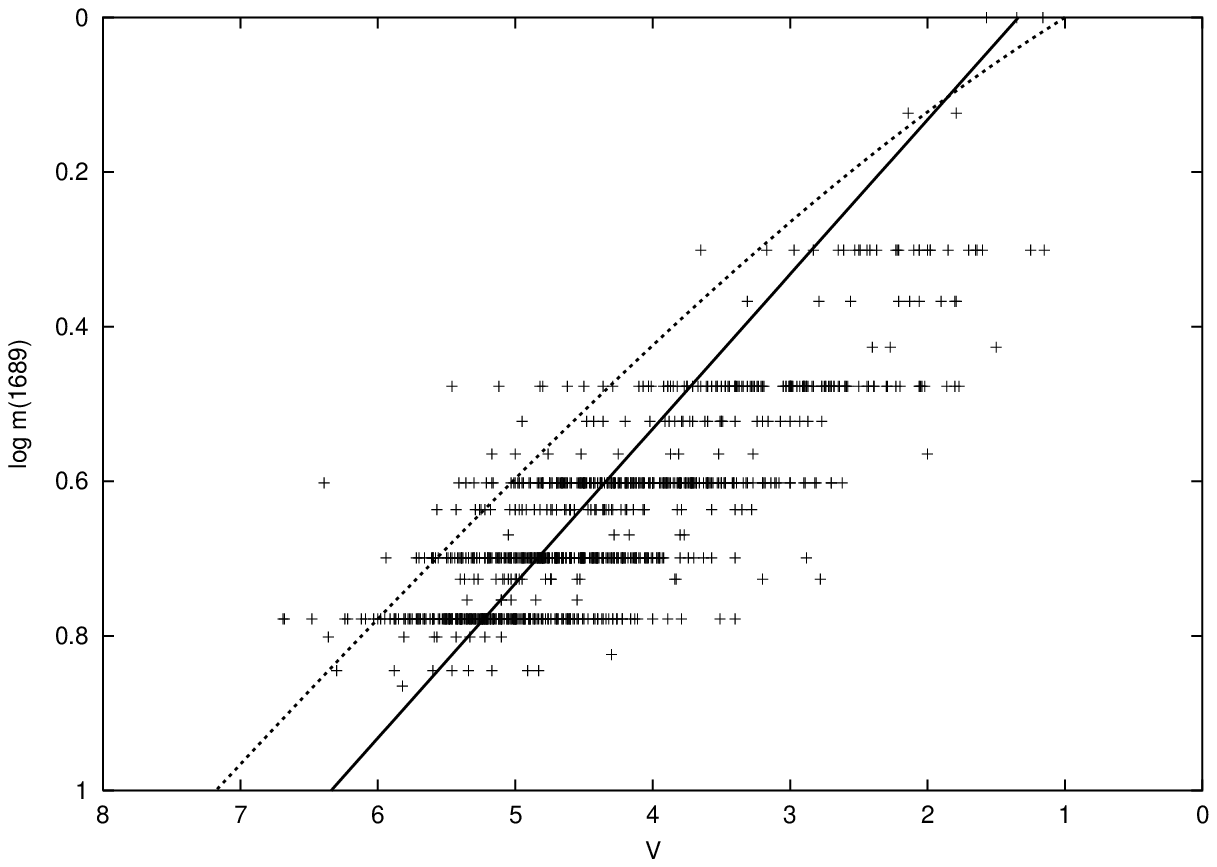}} \\
      \resizebox{60mm}{!}{\includegraphics{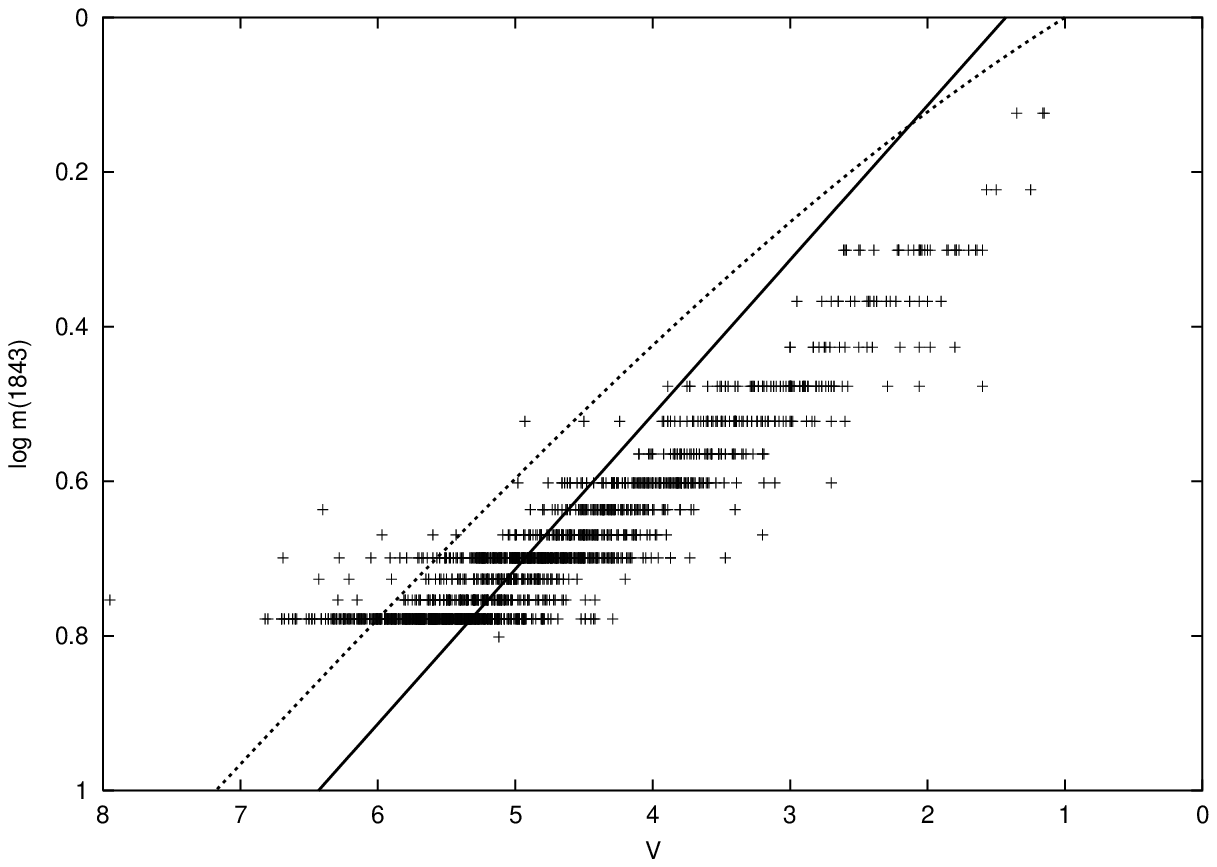}} \\
    \end{tabular}

          \caption{magnitude systems on the power law scale. 
          Dotted lines indicate the function of Schulman \& Cox 
          and solid lines indicate power-law regressions. 
          \label{fig2}}
   \end{figure}

In Sect. 3.1, magnitude systems are found to fit to the logarithmic scale. 
Subsequently, we examine light ratios {\it R} of magnitude systems in old star catalogues. 
In Table~\ref{tbl-3}, we calculate {\it R} in each star catalogue through linear regressions (see Fig.~\ref{fig1}). 
Each {\it R} in old star catalogues approximates to Pogson's $R = 2.512$. 
   
In calculating the value of the light ratio, some data points have to be excluded because of characteristics of records. 
First, with respect to 1st magnitude in old star catalogues, each average of stars recorded as `1' is deviated toward the dimmer magnitude. 
As shown in Sect. 2, we omitted stars recorded as 1st mag considering as unsuitable. 
As well as 1st magnitude in old star catalogues, each average of stars recorded as `6' is deviated toward the brighter magnitude.  
Considering the present ranges of each magnitude, 3rd mag includes stars of 2.5 -- 3.4 mag and 6th mag includes stars of 5.5 -- 6.4 mag.
However, ancient observers defined that the limit of the observable 
magnitude with naked eyes was 6, so their records did not contain 
stars dimmer than 6.0 mag. 
In old star catalogues, 5.5 -- 6.0 magnitude stars were estimated as 6 mag; they recorded only the brightest stars in the range of the 6th mag. 
In `{\it Historia Coelestis Britannica}' (1689), Flamsteed observed and recorded stars of 7th magnitude which was not precisely defined at that time. 
These stars should be considered to be estimated imprecisely. 
In this study, the data of 7th mag is within the purview of 
references and we could not treat them as authoritative data. 

\begin{figure}
   \centering
    \begin{tabular}{cc}
      \resizebox{70mm}{!}{\includegraphics{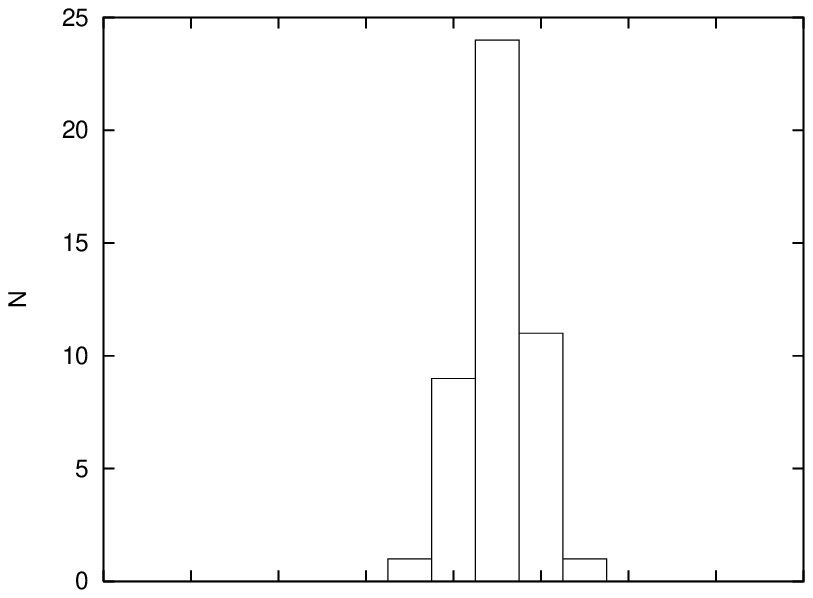}} \\
      \resizebox{70mm}{!}{\includegraphics{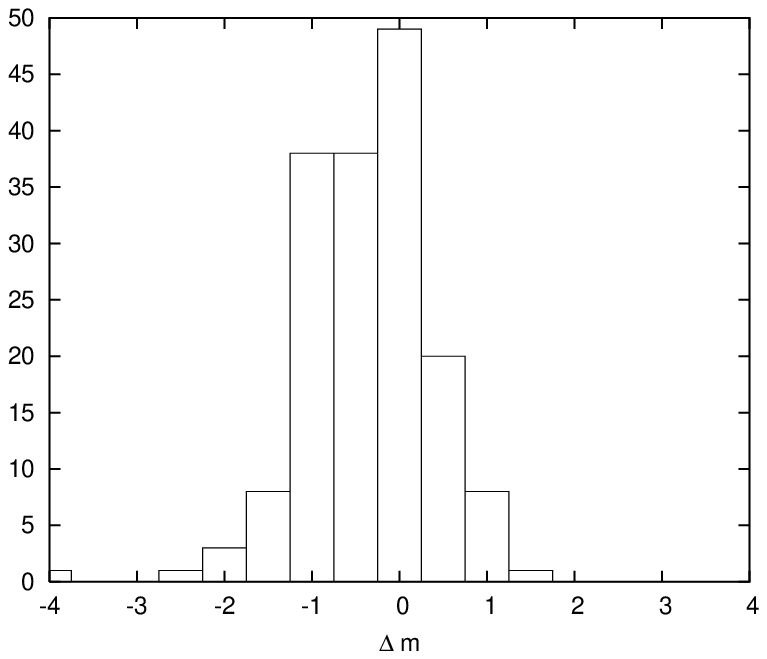}} \\
    \end{tabular}
      \caption{differences of distributions of dispersions between 
      6th (above) and 3rd (below) magnitudes 
         \label{fig3}}
   \end{figure}

In Fig.~\ref{fig3}, we compare distributions of 
dispersion ($m_{137} - V$) in `{\it Almagest}' per recorded 
magnitudes (3rd and 6th).
The distribution of 3rd mag is symmetry around 3.0 mag. 
On the other hand, the distribution curve of 6th mag is not symmetry and found to be cut off. 
Near the observable limit (6.0), there should be recorded stars and non-recorded stars on a fifty-fifty basis. 
Because we could observe and also could not observe stars of observable limit 6th magnitude.
Therefore, the number of the stars of 6.0 mag should be more, and the peak of distribution of 6th mag should be at 6.0. 
As clearly shown in Fig~\ref{fig3}, the distribution of 6th and 3rd magnitude is common for the brighter part, however, for the dimmer part, the distribution of 6th mag is inhibited. 
As shown in Table~\ref{tbl-4}, the standard deviation $\sigma$ of 6th mag in `{\it Almagest}' is much smaller than the others. 
It suggests that the distribution of 6th mag is cut off in half. 
This inhibition should be due to the observable limit with the naked eye, and the criterion of 6th magnitude in old star catalogues should correspond with the same class of Pogson's system. 
As well as 6th magnitude, the distribution of 5th magnitude should be cut off, however, most of this distribution are within the range of magnitudes brighter than 6th.  
Therefore, the part cut off is found to militate hardly for our results. (for detailed values of standard deviations $\sigma$, see Table 3. in Fujiwara et al. 2003)
   
Relative to the magnitude system and the light ratio {\it R} in {\it `Almagest'},  Hearnshaw explained to be logarithmic scale of $R = 3.26$ (1996), and revised $R = 3.42$ (1999). 
These values incline away from Pogson's system. 
Using all magnitudes recorded in {\it `Almagest'}, we confirmed that the light ratio of its system had corresponded to Hearnshaw's value. 
However, we suggest that this difference between Pogson's and Hearnshaw's value should be ascribable to the marginal magnitudes; the brightest magnitude (1st) and the dimmest magnitudes (6th and 7th). 
In order to know real systems of magnitude, we cut inaccurate 1st and 6th mag to draw linear regression. 
As shown in Fig~\ref{fig1} and in Table~\ref{tbl-3}, magnitude systems in old star catalogues, including {\it `Almagest'}, fit to Pogson's scale ($ R = 2.512$).

\section{Conclusions}

   \begin{enumerate}

      \item All magnitude systems in old star catalogues fit to 
      Pogson's logarithmic scale.
      \item On a power-law scale chart, magnitude systems in old star
      catalogues do not have a bias toward proportions at all points, 
      i.e. the power law scale is not consistent with the magnitude 
      systems in old star catalogues.
      \item Relative to 6th magnitude in old star catalogues, mean 
      magnitudes were deviated toward the brighter magnitude due to 
      the range of observable magnitudes. Alike is the 1st magnitude. 
      All linear regressions without these two magnitudes fit to the 
      light ratio $ R = 2.512$ suggested by Pogson. 
   \end{enumerate}

\section*{Acknowledgments}

Fruitful discussions with Dr. P. Kunitzsch at Munich and the hospitality of Observatoire de Paris are greatly appreciated (T.F.). 
We thank to Prof. S. J. Miyoshi for his useful advice on the data analysis.
This work is partly supported by Research Fellowships of the Japan Society for the Promotion of Science for Young Scientists (T.F.), and by a grant-in-aid [14740131 (H.Y.)] from the Japanese Ministry of Education, Culture, Sports, Science and Technology.

\clearpage
\setlength{\paperheight}{200mm}
\setlength{\paperwidth}{135mm}

\begin{table}
\caption{Catalogue ID, observational epoch and number of sampled stars \label{tbl-1}}
\begin{center}
\begin{tabular}{crr}
\hline
ID & epoch & N \\
\hline
1 & 137 & 910 \\   
2 & 964 & 911 \\
3 & 1437 & 889 \\
4 & 1572 & 658 \\
5 & 1603 & 949 \\
6 & 1689 & 1003 \\
7 & 1843 & 1946 \\                     
\hline
\end{tabular}
\end{center}
\end{table}

\begin{table}
\caption{Reduced chi-square in old star catalogues on logarithmic and power-law scale \label{tbl-2}}
\begin{center}
\begin{tabular}{rrll}
\hline
            ID & \multicolumn{1}{c}{{${{\chi}_{\nu}}^{2}$}
            (logarithm)} & 
            \multicolumn{1}{c}{${{\chi}_{\nu}}^{2}$(power-law)} \\            
\hline
            1 & 0.36265 & 1.13955 \\   
            2 & 0.27199 & 0.48512 \\
            3 & 0.28527 & 1.08446 \\
            4 & 0.37404 & 1.01785 \\
            5 & 0.42962 & 1.28275 \\
            6 & 0.38853 & 1.09758 \\
            7 & 0.14192 & 0.92862 \\   
            \hline
\end{tabular}
\end{center}
\end{table}

\begin{table}
\caption{Light ratio {\it R} in old star catalogues \label{tbl-3}}
\begin{center}
\begin{tabular}{rr}
\hline
            ID & R \\
            \hline
            1 & 2.615 \\   
            2 & 2.360 \\
            3 & 2.505 \\
            4 & 2.495 \\
            5 & 2.554 \\
            6 & 2.509 \\
            7 & 2.451 \\
            \hline
\end{tabular}
\end{center}
\end{table}

\begin{table}
\caption{standerd deviation $\sigma$ of each recorded magnitude in `{\it Almagest}' \label{tbl-4}}
\begin{center}
\begin{tabular}{cr}
\hline
            magnitude & $\sigma$ \\
\hline
            1st & 0.91 \\   
            2nd & 0.66 \\
            3rd & 0.73 \\
            4th & 0.56 \\
            5th & 0.51 \\
            6th & 0.39 \\
            \hline
\end{tabular}
\end{center}
\end{table}

\clearpage
\listoffigures

\newpage
%

%

%

\end{document}